\documentclass[letterpaper, 11pt]{article}
 
 \usepackage{lmodern}
 \usepackage{amsmath,amsthm,amssymb,amsfonts}
 \usepackage{hyperref}
 \usepackage{ifthen}
 \usepackage{geometry}
 \usepackage{paralist}
 \usepackage{subcaption}
 \usepackage{appendix}
 \usepackage{graphicx}
 
 \theoremstyle{definition}
 
 \newtheorem{theorem}{Theorem}

 \newtheorem{claim}[theorem]{Claim}
 
 \newtheorem{definition}{Definition}[section]

 \theoremstyle{remark}

\newcommand{\black}[1]{\smallskip \noindent\textbf{#1}}
 
 \usepackage{pifont}
 
\usepackage[charter]{mathdesign}
\usepackage{rotating}
\usepackage{pdflscape} 

\begin{document}
	\title{The Computational Complexity of \emph{Fire Emblem} Series \\and similar Tactical Role-Playing Games}
	\author{Jiawei Gao\thanks{The work was done when the author was a student at University of California, San Diego. Work supported by the Simons Foundation and NSF grant CCF-1909634.}}
	
	\maketitle
	
	\begin{abstract}
		\emph{Fire Emblem (FE)} is a popular turn-based tactical role-playing game (TRPG) series on the Nintendo gaming consoles.
		This paper studies the computational complexity of a simplified version of FE (only floor tiles and wall tiles, the HP and other attributes of characters are constants at most $8$, the movement distance per character each turn is fixed to $6$ tiles), and proves that:
		\begin{enumerate}
		\item Simplified FE is PSPACE-complete (Thus actual FE is at least as hard).
		\item Poly-round FE is NP-complete, even when the map is cycle-free, without healing units, and the weapon durability is a small constant. Poly-round FE is to decide whether the player can win the game in a certain number of rounds that is polynomial to the map size. A map is called cycle-free if its corresponding planar graph is cycle-free.
		\end{enumerate}
		These hardness results also hold for other similar TRPG series, such as \emph{Final Fantasy Tactics}, \emph{Tactics Ogre} and \emph{Disgaea}.
	\end{abstract}
	\nocite{*}
	
	\section{Introduction}
	
	Many video games are proven to be NP-hard or PSPACE-complete \cite{kaye2000minesweeper,friedman2002pushing,biedl2002complexity,demaine2003tetris,cormode2004hardness,hearn2005pspace,forivsek2010computational,fleischer2012algorithmic,johnston2012complexity,viglietta2014gaming,walsh2014candy,viglietta2015lemmings,demaine2016computational,demaine2016super,misra2016two,stephenson2017computational,almanza2018trainyard}. Most of the games are puzzle-solving or path-finding types. However, little research about the hardness of turn-based tactical role-playing games has been conducted.
	
	Tactical role-playing games (TRPGs) are also called ``simulation role-playing games'' (SRPGs) in Japan. Nan et~al.\cite{nan2016turn} calls them ``war chess games''. These games are similar to turn-based board games, where the player and the computer take turns to move their characters in the map. The map is a grid, where characters (which are called \emph{units}) can move from a tile to an empty tile each turn, and can choose to attack a nearby enemy unit or just stay in the tile doing nothing. Each characters has some \emph{attributes}, or \emph{stats}, such as \emph{Hit Points(HP)}, \emph{Attack}, and \emph{Defense}. Usually the characters have personalities and stories like most role-playing games, and are able to level up by getting experience from battles. Well known TRPGs include \emph{Final Fantasy Tactics}\cite{fft}, \emph{Tactics Ogre}, and \emph{Disgaea}\cite{disgaea}.
	
	\emph{Fire Emblem(FE)}\cite{USofficial,JPofficial} is a Japanese TRPG series developed by INTELLIGENT SYSTEMS and released on Nintendo gaming consoles and handheld gaming devices. It has 29 years of history, with 16 titles in its main series.
	The feature of FE is that all numbers involved in battles (such as attack, defense, and damage) are relatively small (compared to most other popular TRPGs, e.g.,\,\emph{Final Fantasy Tactics}) and completely transparent to the player. The mechanism of the game is well-studied, and the enemy units' strategy can be predicted by experienced players. The stages of the game are so challenging that it encourages advanced-level players to precisely calculate the damages dealt and taken, and also predict or even manipulate the enemy units' moves. Another uniqueness of the series is that in battles, the defending units can counter attack, so in many cases it is wise no to approach the enemies rashly, but to form a defensive formation and wait for the enemies to approach and get defeated by counter attacks. In most other TRPGs, however, not all units can counter attack automatically, although some units have counter attack as their skills.
	
	In each turn, each unit is granted to move once.
	A unit can move from zero to \emph{Mov} tiles, where \emph{Mov} is an attribute of the unit indicating how far they can move in a single turn. All distances in the game are Manhattan distance, or in other words, computed by the L$1$-Norm: the distance from coordinate $(0,0)$ to $(0,1)$ or $(1,0)$ is only $1$, but to coordinate $(1,1)$ would be $2$.
	
	If there is a hostile unit within the attack range of a unit (again by Manhattan distance), then the unit can initiate a combat.
	When combat happens, the amount of damage the defender takes from the attacker is computed by the following formula:
	\[
	\text{damage to the defender} = \begin{cases}
	\text{attacker's }\mathit{Atk} - \text{defender's }\mathit{Def}\text{, if hit}\\
	0\text{, if miss}
	\end{cases}
	\]
	After combat, the HP of the defender is decreased by the amount of the damage.
	A unit dies if their HP is below zero. If the defender is still alive, they can counter attack the attacker, using the same damage formula with their roles switched.
	
\begin{figure}
		\centering
		\begin{subfigure}[b]{0.5\linewidth}
			\includegraphics[width = \linewidth]{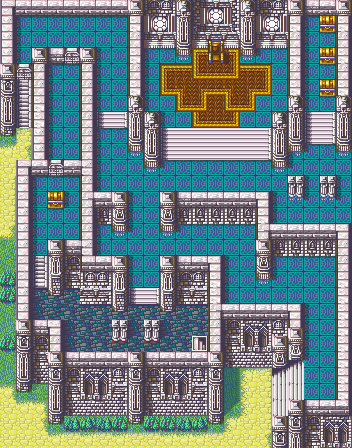}
			\caption{A map of a stage}
			\label{fig_intro_stage}
		\end{subfigure}
		\begin{subfigure}[b]{0.4\linewidth}
			\includegraphics[width = \linewidth]{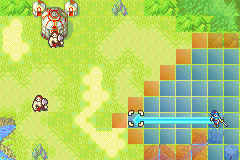}
			\caption{A unit moves on the map. The blue tiles are positions she can move to, and the orange tiles are positions she can attack.}
			\label{fig_intro_move}
			\includegraphics[width = \linewidth]{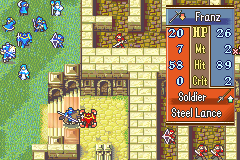}
			\caption{A unit (Franz) attacks an enemy (Soldier).}
			\label{fig_intro_atk}
		\end{subfigure}
		\caption{Screenshots of the game.}
		\label{fig_intro}
	\end{figure}

	Figure~\ref{fig_intro_stage} shows the map of a stage.
	Figure~\ref{fig_intro_move} shows a gameplay screenshot of a player unit about to move. Figure~\ref{fig_intro_atk} shows a screenshot of a combat.
	
	The details of the game mechanism used in this paper will be presented in Section~\ref{sec_pre}.
	
	\section{Main Results}
	
	The actual game play of FE series have many complicated mechanisms. Because it involves a lot of numbers and computing, proving the hardness of a game where the HP of units is unbounded would not be an interesting result. Therefore we consider the simplified FE game. In the games, there are only floor tiles and wall tiles, the HP and other attributes of characters are constants at most $8$, the movement distance per character each turn is fixed to $6$ tiles. This is a special case of real FE game, and can be easily adapted to most other similar TRPGs.
	
	Because the games studied in this paper may take too many turns than actual gameplay, we do not consider weapon or healing staff durability, or equivalently, let the durability of all weapons and healing staves be infinity.
	
	\begin{definition}[Simplified FE]
		Given the initial status of an FE game with infinite weapon and staff durability,
		decide whether the player has a winning strategy.
	\end{definition}
	
	This paper proves the hardness of Simplified FE:
	
	\begin{theorem}\label{thm_pspace}
		Simplified FE is PSPACE-complete.
	\end{theorem}
	
	Even if it is not surprising that a two-player turn-based strategy game is PSPACE-complete, FE is not as complicated as a typical two-player turn-based strategy game, because the enemies' strategy is simple, deterministic\footnote{In real FE, the outcome after each combat is not deterministic, because the amount of damage is randomized, affected by the rate of missing and the rate of critical hits. But the strategy of where each enemy unit moves to and which player unit they attacks, is always deterministic.}, and can be fully predicted based on the current game configuration. Each enemy unit acts like a robot pre-programmed with simple and straightforward commands, like ``move to your target by the shortest path, and hit the target''. 
	For example, if an enemy unit can reach a player unit and attack them, then the enemy unit will surely choose to do that, even in the case where the enemy unit can make little damage to the player unit, and would die from the player unit's counter attack.
	Another example is that, if two enemy units both attack the player's Lord, the Lord would be killed by the total damage and the game would be lost. But if we place a weaker Cleric unit near to the Lord, one of the enemy units would attack the Cleric, and the other enemy unit would attack the Lord, leaving both the Cleric and the Lord alive.
	
	Thus, instead of being a game of form ``$\exists$ a player strategy, $\forall$ enemy strategies, $\exists$ a player strategy, $\forall$ enemy strategies ...'' which can be PSPACE-complete even when the total number of turns is bounded by a polynomial, the game is more similar to the case ``$\exists$ a player strategy, for \emph{the} enemy strategy, $\exists$ a player strategy, for \emph{the} enemy strategy, ...''. Therefore, the result that FE being PSPACE-complete is not trivial.
	
	The reason of FE being PSPACE-complete is that it may take too many turns.
	Actually, if one can make sure that a game ends in a polynomial number of turns, then the game is in NP.
	
	In reality, FE stages usually do not take too many turns, observing the following facts:
	
	(1) After each turn, the total HP of all units is decreasing fast from the battles, even if some healing happens occasionally.
	
	(2) In most cases, the enemy units usually move towards player units, and do not move away. Even if some enemy units do not move if player units do not get near them, there is little point of waiting a lot of turns before attacking.
	
	(3) Weapons and healing staves have durability that are usually around $40$.
	
	(4) Sometimes the game design encourages or requires players to clear each stage as fast as possible.
	
	So it would be interesting to ask if a stage can be finished in a certain number of rounds.
	
	\begin{definition}[Poly-round FE]
		Given the initial status of an FE game and a number $k$, where $k$ is represented in unary,
		decide whether the player has a winning strategy with at most $k$ rounds.
	\end{definition}
	
	Poly-round FE is in NP, because a winning strategy always has a polynomial number of turns.
	
	Cycles in the map (e.g.\,a closed path surrounding a stone pillar, or a plain with a lake in the middle) usually make stages more complicated. For example, if there are two paths leading to the same room, we can lure enemies to one path and go through the other. Player unit can move along a cycle for many rounds so that the enemy chasing them will never reach them; this strategy is useful when we need a lot of turns to heal the units, or to wait for powerful allies in different areas of the map to come and deal with the enemies.
	Also, we can let a player unit move back and forth at one side of the cycle, making an enemy moving back and forth at the opposite side of the cycle, because the shortest path for the enemy unit to reach the player unit keeps changing.
	
	Thus, the game would be much simpler if we do not allow cycles in the maps.
	
	\begin{definition}[Cycle-free FE]
		If a map does not have cycles when considered as a planar graph where floor tiles are vertices and adjacent floor tiles are connected by edges, then the stage is called \emph{cycle-free}.
	\end{definition}
	
	This paper shows that even if the stage is cycle-free, Poly-round FE is still NP-complete. Because the construction in the NP-completeness proof does not involve too many combats, we can limit the weapon durability to a small constant. Also, in the construction, there are no healing units.
	
	\begin{theorem}\label{thm_np}
		Poly-round FE is NP-complete, even if the map is cycle-free, the weapon durability is a constant at least $4$, and there are no healing units.
	\end{theorem}
	
	These hardness results can also be applied to other TRPGs with similar mechanism, although in those TRPGs, units cannot counter attack when being attacked.
	
	\section{Game Mechanism}\label{sec_pre}
	
	In this section we will describe our simplified FE model used in the proofs. This simplified model fits into most real FE games, thus the hardness of these simplified cases can demonstrate the hardness of real FE games.
	
	\black{Stage}
	
	The map of a stage is an $m \times n$ grid, where each tile is a square adjacent to four other tiles in its four directions. A tile can either be castle floor or castle wall. Units can be placed in the floor tiles, but cannot move onto or through walls.
	However, ranged units (e.g.,\,archers who have attack range $2$) can attack through walls (imagine there are little holes in the walls for arrows to fly through). The distance between two tiles is defined to be their Manhattan distance.
	
	\black{Units}
	
	A unit is a character, which occupies one tile in the map. A unit can either be a player unit, or an enemy unit. Some units, such as Clerics, can heal their allies. Each unit has the following attributes:
	\begin{compactitem}
		\item\emph{HP:} the Hit Point of the unit, an integer between $1$ and $50$.
		\item\emph{Atk:} the Attack of the unit, an integer between $1$ and $50$.
		\item\emph{Def:} the Defense of the unit, an integer between $1$ and $50$.
		\item\emph{Mov:} the maximum Manhattan distance the unit can move each turn, an integer at most $10$.
		\item\emph{Range:} the attack range of the unit. If an enemy's distance is in the unit's attack range, then the unit can attack the enemy. The most common attack range of a unit is usually either one or two.\footnote{Unlike real FE where the attack range is affected by the weapons equipped by the units, here we assume each unit has a fixed attack range.} Healing units have healing range, where an ally in the range can be healed by the unit.
	\end{compactitem}
	
	Units of the same class always have the same \emph{Mov} value.
	The most common \emph{Mov} value is $6$, which is the movability of infantry units. In the proofs, we will always make all units have the same \emph{Mov} and use letter $d$ to represent this \emph{Mov} value.
	
	If a unit's HP is zero or below, the unit is dead, and disappears immediately from the map, and will not appear again anywhere.
	
	\black{Moving}
	
	In each round, the player's turn comes first, where the player can move all the player units; then comes the enemies' turn, where each enemy unit moves, controlled by the computer. The player can move the player units in any order. In a single round, the order of the player units' moves and the enemy units' moves do not interleave.
	
	In a turn, each unit can move at most one time. A unit can move to an empty tile along a valid path within distance of their \emph{Mov}. They can also stay in their own tile without moving. After moving to the target tile, if an enemy unit is in their attack range, then they can choose to attack the enemy unit. Or they can choose to heal an ally, if there are allies in their healing range.
	
	Units can move through empty floor tiles, or floor tiles with other units of their own allies. But they cannot move through wall tiles or floor tiles occupied by hostile units.
	
	\black{Combat}
	
	If attacker $a$ hits defender $d$, then $d$ takes damage equal to ($a$'s \emph{Atk} $- d$'s \emph{Def}). $d$'s HP is decreased by the damage value.
	If $d$ is still alive (HP $> 0$) and $a$ is in $d$'s attack range, then $d$ immediately counter attacks $a$. $a$ takes damage equal to ($d$'s \emph{Atk}$- a$'s \emph{Def}). Afterwards, $a$'s HP is decreased by the damage value.
	
	In the hardness proofs, we only consider the cases where attacks always hit and never miss.
	
	\black{Objective}
	
	The most common objective of FE stages is to ``seize the throne''. There is a unique tile in the map representing a throne, and there is a main character among the player units called the Lord. The goal of the stage is to move the Lord to the throne tile (if it is not occupied by enemies, of course.)
	If the Lord is dead, then the player loses the game immediately.
	
	In real FE stages, there are also other objectives than seizing the throne, such as rout the enemy, defeat the enemy leader, survive for a certain number of turns or escape from the exit. This paper will not use these objectives in the hardness proofs.
	
	\black{Enemy strategies}
	
	In FE, there are two types of enemy units:
	
	\begin{itemize}
		
		\item \emph{Impatient enemy:} If there are player units in their move-and-attack range (i.e.,\,if the enemy moves, then the player unit is in the enemy's range), then they will choose the player unit with the lowest defense, move to the unit by the shortest path and attack the unit. Otherwise, if there are player units in the same connected component of the map, then they will choose the player unit in the same connected component with the lowest defense as their target, and moves to the target by the shortest path (even if the path may be blocked by another player or enemy unit).
		
		\item \emph{Patient enemy:} If there are player units in their move-and-attack range, then they will choose the player unit with lowest defense, move to the unit by the shortest path and attack the unit. Otherwise, they do not move at all. Once the patient enemy takes part in any combat, they will turn into an impatient enemy and will not turn back to patient ones ever again.
	\end{itemize}
	
	The order of enemy units’ moves is fixed, by the implicit order that enemy units are created
	on the map. In the hardness proofs we will make the stages order-invariant.
	If there are multiple targets for an enemy, they will break ties by the order that the player units are created.
	If there are multiple shortest paths, they will break ties by selecting a path that moves horizontally as far as possible and then moves vertically. In the hardness proofs we will avoid creating multiple targets or multiple paths for enemies.
	
	In this paper, for simplicity we consider the game setting where there are no critical hits, dodged or ineffective attacks, personal skills, special effect weapons, terrain effects, support effects, item exchange, and weapon triangles, even if these features appear in many FE titles. Also we assume characters will not level up in the stage, thus the values of attributes of a unit will never change. For readers who are FE players, consider the case where all unrelated attributes like \emph{speed}, \emph{skill}, \emph{luck} are zero, all units use the same type of weapon, and the experience gained by combats is too small to level up units. The proofs will avoid the possibility for terrain effects, support effects and item exchange.

	\section{The NP-hardness of Poly-round Cycle-free FE}\label{sec_nph}
	
	This section proves Theorem~\ref{thm_np}.
	We reduce from Rectilinear Monotone 3-Bounded 3-SAT \footnote{\cite{darmann2016planar} and \cite{tippenhauer2016planar} used the term "Planar Monotone 3-Bounded 3SAT" to define a different problem, where the planar graph does not require positive clauses and negative clauses to be placed on opposite sides. So here we use the word "Rectilinear" as in \cite{de2010optimal}, even if \cite{de2010optimal} does not use "Rectilinear" as part of the problem name.}. 
	It is a 3-SAT problem satisfying the following conditions:
	\begin{compactitem}
		\item In each clause, either all literals are positive, or all literals are negative.
		\item Each variables appears as a positive literal in at most three clauses, and as a negative literal in at most three clauses.
		\item If we treat each variable and each clause as a vertex, and connect a pair of variable and clause by an edge if the variable appears in the clause, then the resulting graph is a planar graph.
		\item Moreover, in the planar graph, we can place all variable vertices in a line, put all clauses with positive literals on one side, and put the clauses with negative literals on the other side.
	\end{compactitem}
	
\begin{figure}
		\centering
		\includegraphics[scale=0.9]{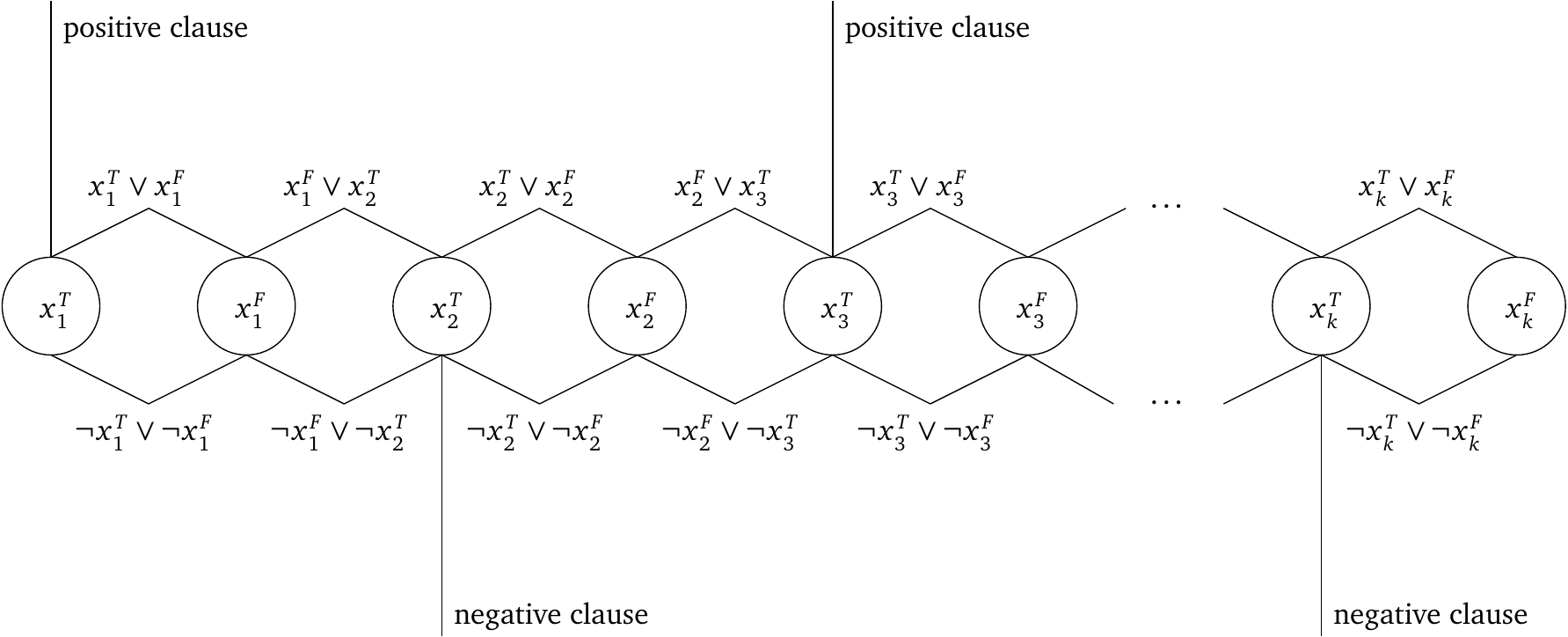}
		\caption{Reduction from Planar Monotone 3-SAT to Rectilinear Monotone 3-Bounded 3-SAT.}\label{fig_sat}
	\end{figure}

	To show Rectilinear Monotone 3-Bounded 3-SAT is NP-complete, we reduce from Planar Monotone 3-SAT \cite{de2010optimal}, which is NP-complete. This problem is similar to Rectilinear Monotone 3-Bounded 3-SAT, but the number of clauses each variable appears in is unbounded. To make each variable appear in at most three positive clauses and three negative clauses, we make the following modification on each variable:
	
	For variable $x$ that appears in $k$ clauses, we create new variables $x_1^T, x_1^F, \dots, x_k^T, x_k^F$, and connect them by clauses as shown in Figure~\ref{fig_sat}. In this way, $x_i^T$ and $x_i^F$ must have different truth values, and $x_i^F$ and $x_{i+1}^T$ must have different truth values. So either all $x_i^T$ are true and all $x_i^F$ are false, or all $x_i^T$ are false and all $x_i^F$ are true. Because all $x_i^T$ have the same truth value, we will let these variables substitute all the appearances of variable $x$ in clauses. If $x$ is in some positive clause, we connect it to the clause on one side. If it is in some negative clause, we connect it to the clause on the other side.
	
	\bigskip
	Next we will reduce Rectilinear Monotone 3-Bounded 3-SAT to Poly-round Cycle-free FE.
	For each variable, we create a Player Variable Unit. For each clause, we create a Player Clause Unit. And we create another player unit which is the Lord, whose objective is to seize the throne.
	For each literal we create an Enemy Literal Unit. For each variable, we create $2d-1$ Enemy Sniper \footnote{A Sniper is a promoted Archer. This paper calls them Snipers instead of Archers because they are so dangerous that can easily defeat a player unit.} Units, where $d$ is the \emph{Mov} of the units. In the figures, $d$ is shown to be $6$.
	Table \ref{table_np} shows the attributes of all units in the construction.
	
	\begin{table}
		\centering
		\begin{tabular}{|l|c|c|c|c|}
			\hline
			Unit & HP & Atk & Def & Range\\
			\hline
			Player Variable Unit & 4 & 3 & 1 & 2\\
			Player Clause Unit & 3 & 2 & 1 & 2\\
			Player Lord & 1 & 2 & 1 & 1\\
			\hline
			Enemy Literal Unit & 2 & 2 & 1 & 2\\
			Enemy Sniper Unit & 2 & 5 & 2 & 2\\
			\hline
		\end{tabular}
		\caption{Attributes of all units in the NP-hardness proof. All units has Mov = 6.}
		\label{table_np}
	\end{table}
	
	\black{Variable gadget}
	
\begin{figure}
		\centering
		\includegraphics[width = .9\textwidth]{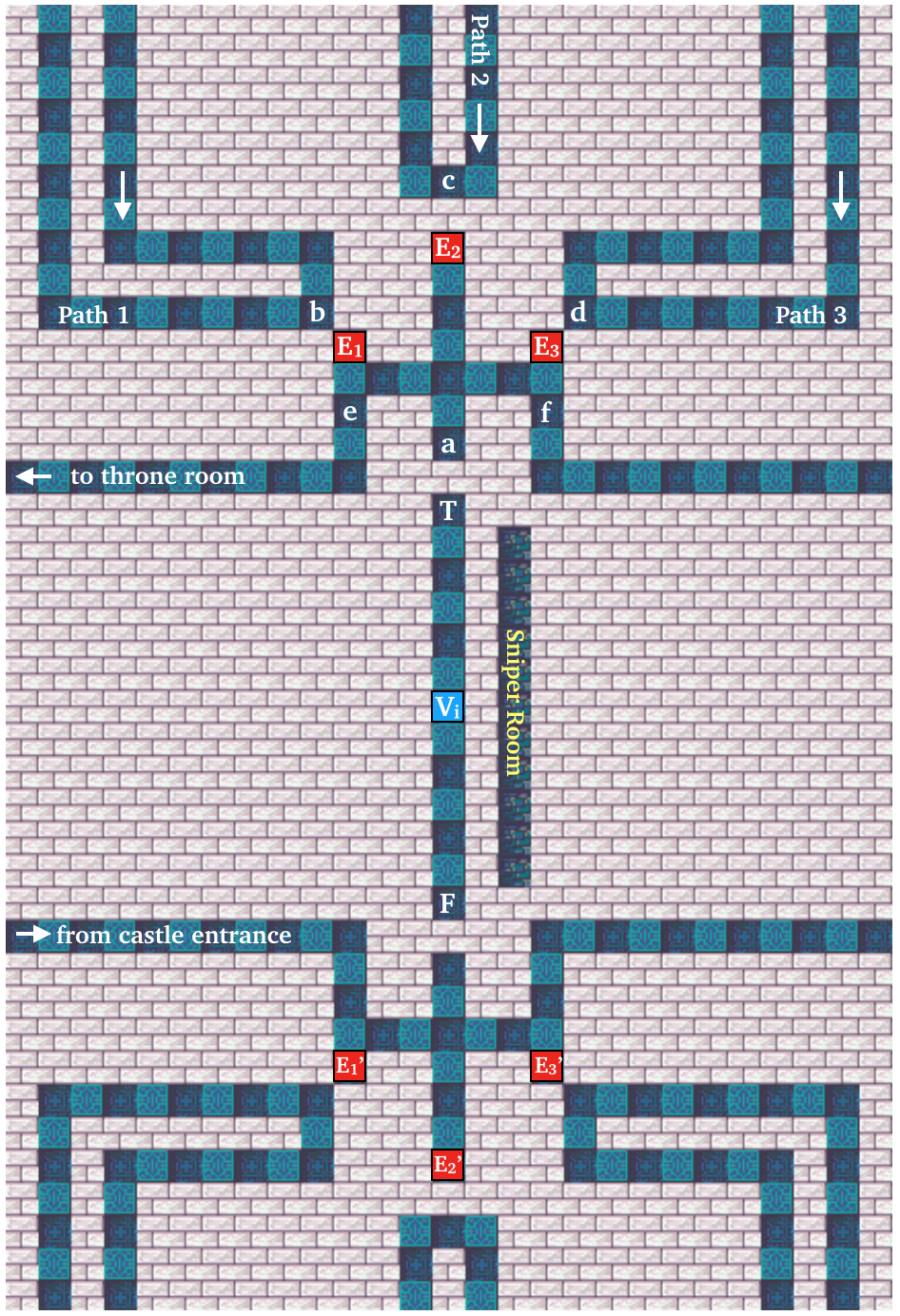}
		\caption{Variable gadget. The blue carpet tiles represent the floor. The lighter and darker blue tiles are the same, where the colors are aesthetic design to show the tiles clearly, and do not have particular meaning. The brick tiles represent castle walls.}\label{fig_np_var}
	\end{figure}

	Figure \ref{fig_np_var} shows a variable gadget. In the center of the $i$-th variable gadget we put a Player Variable Unit $V_i$. $V_i$ is located in a vertical long room of distance $2d+1$. Moving upwards corresponds to setting variable $v_i$ to true, while moving downwards corresponds to setting it to false.
	
	Above $V_i$, there are at most three patient enemy units $E_1, E_2, E_3$. Each enemy unit corresponds to a positive literal of the variable $v_i$ in some clause. Similarly, below $V_i$, there are also at most three enemy units $E_1', E_2', E_3'$, each corresponding to a negative literal of $v_i$ in some clause.
	
	If $V_i$ moves up to the tile labeled $T$, then all three enemy units above will be alert because $V_i$ is in their range, because their distance to the tile labeled $a$ is $d$. They will move down to position $a$ to attack $V_i$. Each enemy unit makes damage $1$ to $V_i$, so $V_i$ will lost at most $3$ HP. In the counter attacks, $V_i$ makes damage $2$ to each of the enemy units, so the three enemy units will not survive $V_i$'s counter attacks. Thus, if $V_i$ moves up to position $T$, the above three enemy units will disappear from the map; similarly, if $V_i$ moves down to the tile labeled by $F$, the three enemy units below will disappear.
	
	The room labeled ``sniper room'' to the right of $V_i$'s room has length $2d-1$ and is filled with Enemy Sniper Units of range $2$, with \emph{Atk} so high that one single shot will take $V_i$'s life, and \emph{HP} + \emph{Def} high enough so that they cannot be defeated by $V_i$ by one shot. Thus, $V_i$ must move to either position $T$ or position $F$ in the first turn. Once $V_i$ has chosen to move up to position $T$ in the first turn, then $V_i$ cannot move down to position $F$, and vice versa.
	
	Above the three enemy units there are three paths. The three paths wind their way around the three enemy units, with the nearest distance to the enemy units being $2$. The nearest tiles are labeled by $b$, $c$ and $d$. These paths correspond to the clauses the variable $v_i$ is contained in. Units in clause gadgets will go along these paths to encounter the enemy units.
	
	\black{Clause gadget}
	
	We consider the clauses with exactly three literals. A gadget for clause $c_j$ is a long path which goes around the three enemy units corresponding to its three literals appearing in clause $c_j$. The paths for clauses with all positive literals are placed above the variable gadgets, and the clauses with all negative literals are below them. A Player Clause Unit $C_j$ needs go along the path to clear all the remaining Enemy Literal Units in their own clause, that are not cleared by any Player Variable Units. The path itself is a connected component, so $C_j$ cannot move to any variable gadget or other clause gadgets. Because the variable-clause graph corresponding to the CNF formula is a planar graph, the clause gadgets can be embedded in the grid map without intersecting each other.
	
	If $C_j$ is at location $b$ (or $c,d$), then $C_j$ a chance to attack enemy unit $E_1$ (or $E_2, E_3$ respectively).
	The battle between $C_j$ and an enemy takes two turns:
	
	\begin{compactenum}
		\item Enemies' turn: Enemy unit damages $1$ HP to $C_j$, $C_j$ damages $1$ HP to enemy unit.
		\item Player's turn: $C_j$ damages $1$ HP to enemy unit. Enemy unit is dead and cannot counter attack.
	\end{compactenum}
	
	This is the best strategy for $C_j$. For otherwise if in a turn, $C_j$ attacks first, then it will be damaged by $1$ HP from the counter attack. Then in the enemies' turn, the enemy unit will kill $C_j$.
	
	Thus, $C_j$ loses at least $1$ HP when taking out an enemy unit. $C_j$ has $3$ HP in the beginning, thus if there are three enemy units remaining in the clause (which means all three literals are false), then $C_j$ not able to defeat all of them. Otherwise, $C_j$ can clear all enemy units corresponding to their clause.
	
	For clauses with fewer than $3$ literals, we modify the corresponding $HP$ of $C_j$ to be the number of literals in the clause.
	
	\black{Main Road}
	
	The player Lord unit enters the castle from bottom left gate, and goes along the main road inside the castle which leads to the throne. The main road traverses through the negative sides of all variable gadgets, and then through the positive sides of all variable gadgets, and finally reaches the throne. The Lord is very weak: she has attack range only one, and cannot survive any attack from an enemy unit. The lord can only pass through all variable gadgets if all Enemy Literal Units are cleared.
	
	In Figure~\ref{fig_np_var}, suppose the Lord is moving through the positive part of the variable gadget, from the right of the map to the left of the map. The road goes through the trident shaped room where the enemy units $E_1, E_2$ and $E_3$ are located. If the Lord moves to a tile between the tiles labeled by $f$ and the tile labeled by $e$, including $f$ and $e$, then she will be attacked by all three enemy units $E_1, E_2, E_3$. Also the Lord cannot avoid ending a turn in tiles between $e$ and $f$, because the distance from $e$ to $f$ is greater than $d-1$. Thus, the Lord can only pass when all three enemy units have been cleared: which means either they have been defeated by the variable unit (the variables is set to true so that the literals are true) or they have been defeated by the clause unit (the clause value is true even if these literals are false).
	
\begin{landscape}
		\begin{figure}[t]
			\includegraphics[height = \textwidth]{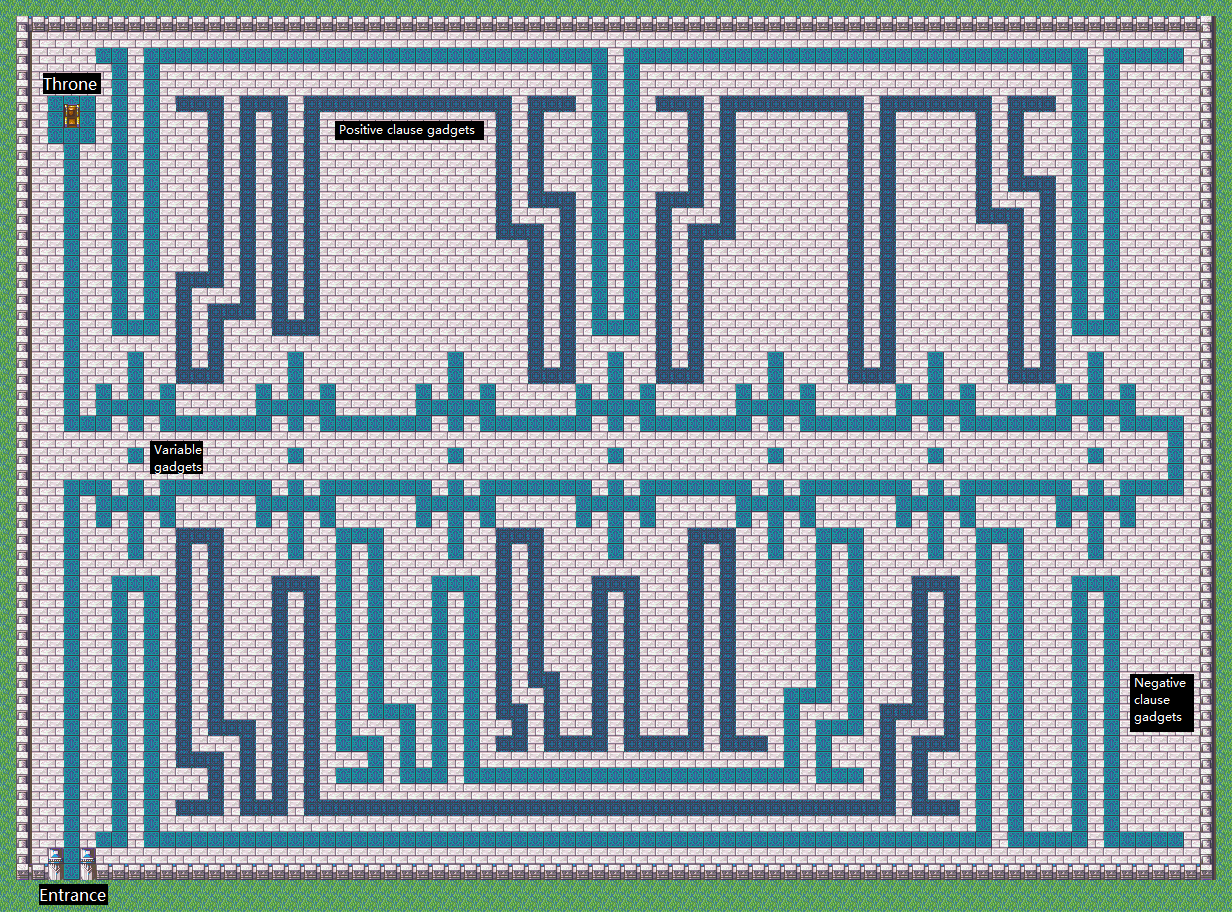}
			\caption{Overview of the map showing simplified variable gadgets, clause gadgets and the main road. The lighter and darker blue tiles are the same, and the colors do not have particular meaning.}\label{fig_np_overview}
		\end{figure}
	\end{landscape}

	Figure \ref{fig_np_overview} shows a simplified overview of the whole map. In this example, the 3-CNF formula is 
	\begin{multline*}\varphi = (v_1 \vee v_2 \vee v_4) \wedge
	(v_4 \vee v_6 \vee v_7) \wedge
	(v_1 \vee v_4 \vee v_7) \wedge\\
	(\neg v_1 \vee \neg v_6 \vee \neg v_7) \wedge
	(\neg v_1 \vee \neg v_2 \vee \neg v_6) \wedge
	(\neg v_2 \vee \neg v_3 \vee \neg v_5) \wedge
	(\neg v_3 \vee \neg v_4 \vee \neg v_5)
	\end{multline*}

	If there is a satisfying assignment to the CNF-formula, then in each clause at least one Enemy Literal Unit is cleared by the move of a Player Variable Unit, then the Player Clause Unit can clear all the remaining Enemy Literal Units, and thus Lord can always reach the throne.
	In the other direction, if the Lord can reach the throne, then it means all Enemy Literal Units are cleared, so each clause has at least a true literal, thus the Player Variable Units' moves gives a satisfying assignment of the CNF formula. Thus Theorem~\ref{thm_np} follows.
	
	Moving all the $V_i$ takes one round. After the first round, all $C_j$ moves in their paths in one direction, and the Lord moves along the main road in one direction. Thus the total number of rounds taken to win is bounded by the total number of tiles of the map.
	
	In the construction, the total number of attacks and counter attacks by each unit is at most $4$ times, so it works even when the weapon durability in the game is a fixed constant greater than or equal to $4$.
	
	\bigskip
	
	For other TRPGs where units cannot make counter attacks, we can adapt our construction for these games. The new table of attributes is Table~\ref{table_np2}, where modified values are shown in bold font.
	
	\begin{table}
		\centering
		\begin{tabular}{|l|c|c|c|c|}
			\hline
			Unit & HP & Atk & Def & Range\\
			\hline
			Player Variable Unit & \textbf{7} & 3 & 1 & 2\\
			Player Clause Unit & 3 & 2 & 1 & 2\\
			Player Lord & 1 & 2 & 1 & 1\\
			\hline
			Enemy Literal Unit & 2 & 2 & 1 & 2\\
			Enemy Sniper Unit & 2 & \textbf{8} & 2 & 2\\
			\hline
		\end{tabular}
		\caption{Attributes of all units in the NP-hardness proof, in TRPGs without counter attack.}
		\label{table_np2}
	\end{table}
	
	Variable Gadget: When $V_i$ moves to position $T$ (or $F$), they will be attacked at most three times by $E_1, E_2, E_3$ (or $E_1', E_2', E_3'$), losing at most $3$ HP. In the next round, $V_i$ kills one enemy unit, and takes at most $2$ damage. In the third round, $V_i$ kills another enemy unit, taking at most $1$ damage from the last enemy unit. Finally, $V_i$ kills the last enemy unit. So $V_i$ needs to have at least $7$ HP in the beginning. The Enemy Snipers need to have \emph{Atk} at least $8$ to be able to kill $V_i$ by a single shot.
	
	Clause Gadget: To defeat an Enemy Literal Unit, $C_j$ must lose at least one $1$ HP. The way for $C_j$ to defeat an Enemy Literal Unit $E$ losing only $1$ HP is as follows:
	\begin{compactenum}
		\item Player's turn: $C_j$ damages $1$ HP to enemy unit.
		\item Enemies' turn: Enemy unit damages $1$ HP to $C_j$.
		\item Player's turn: $C_j$ damages $1$ HP to enemy unit. Enemy unit is dead.
	\end{compactenum}
	
	In the TRPGs where the goal of a stage is to defeat all enemies or defeat the enemy leader, we could replace the throne by a patient enemy leader with \emph{Atk}, \emph{HP} and \emph{Def} being one, so that they can be defeated by the Lord.
	
	Thus, this reduction works for these TRPGs, as long as they support distance $2$ attacks and patient enemies.

	\section{Simplified FE is PSPACE-complete}\label{sec_pspace}
	
	This section proves Theorem~\ref{thm_pspace}: Simplified FE is PSPACE-complete.
	We use the technique introduced in \cite{viglietta2014gaming} and later used in \cite{viglietta2015lemmings, aloupis2015classic, demaine2016super}, which is called the ``open-close door'' framework:
	
	\emph{In a game whose goal is to move a player-controlled avatar from the starting location to an ending location along a path, if we can create \emph{open-close door gadgets} and \emph{crossover gadgets}, then there is a reduction from True Quantified Boolean Formula (TQBF) to the game.}
	
	An \emph{open-close door gadget} has a door with three states, an initial state (which may be the same as the close state), an open state and a close state.
	
	There are three distinct paths going through the gadget: a traverse path, an open path, and a close path.
	\begin{compactitem}
		\item The avatar can go through the traverse path in any direction if and only if the door is in the open state.
		\item If the avatar goes through the open path, they may push an ``open'' button to let the door open.
		\item If the avatar goes through the close path, they are forced to push a ``close'' button to let the door close.
	\end{compactitem}
	
	The door should be able to be opened, closed, and travered through for an arbitrary number of times.
	
	Though not explicitly stated in the papers, their construction actually has the following fact:
	\begin{claim}\label{claim_oneway}
		This framework works even if the paths through the doors are not strictly one directional, allowing the avatar to traverse the paths reversely, and
		\begin{compactitem}
			\item When the close path is traversed reversely, the door is forced to open;
			\item When the open path is traversed reversely, the door is forced to close;
			\item When open path is traversed in the correct direction, the door is forced to open.
			\item When the door is already open and the open path is traversed again in the correct direction, the door remains open;
			\item When the door is already closed and the close path is traversed again in the correct direction, the door remains closed.
		\end{compactitem}
	\end{claim}
	The last three points of the above claim can be observed from the quantifier gadgets and clause gadgets constructed in \cite{viglietta2014gaming}. The first two points will be proved in Section~\ref{sec_oneway}.
	
	A \emph{crossover gadget} is a gadget to enforce any two paths A and B cross each other without leakage. That is, an avatar on A and should continue on path A and should not end up on path B, and vice versa.

	\bigskip
	
	We construct a FE stage which supports the open-close door gadgets and the crossover gadgets.
	In our construction, the player has one main unit, which we call the Lord. We will also create a lot of healing units, who can heal at least $2$ HP to the Lord each time. On the enemy side, there is a Dragon, corresponding to each door gadget there is a Door Gadget Unit, and also there are a lot of Sniper Units that will be useful in the crossover gadgets.
	Table \ref{table_pspace} shows the attributes of all units in the construction. All units has \emph{Mov} value $d = 6$.
	\begin{table}
		\centering
		\begin{tabular}{|l|c|c|c|c|}
			\hline
			Unit & HP & Atk & Def & Range\\
			\hline
			Player Lord & 3 & 1 & 1 & 2\\
			\hline
			Enemy Dragon & 5 & 5 & 5 & 1\\
			Enemy Door Gadget Unit & 2 & 2 & 2 & 2\\
			Enemy Sniper& 5 & 5 & 5 & 2\\
			\hline
		\end{tabular}
		\caption{Attributes of all units in the PSPACE-completeness proof. }
		\label{table_pspace}
	\end{table}
	
	In the construction, our Lord has attack range $2$. Her goal is to travel from her starting position to the throne.
	There is a narrow path and the Lord has to run at full speed without stopping. A dangerous dragon $D$, who is an impatient enemy with attack range $1$, is placed behind her initially at distance $2$. The Lord must move distance $d$ along the path each turn, for otherwise $D$ will reach her and kill her. Every time the player's turn ends, the dragon moves to a tile behind the Lord, keeping distance $2$. 
	Thus, the Lord is forced to end her turns in the tiles of distance $i$ along the path where $i \mbox{ mod } d = 0$. We call these tiles the \emph{safe tiles}.
	
	\black{Door gadget}
	
	\begin{figure}
		\centering
		\includegraphics[scale=0.6]{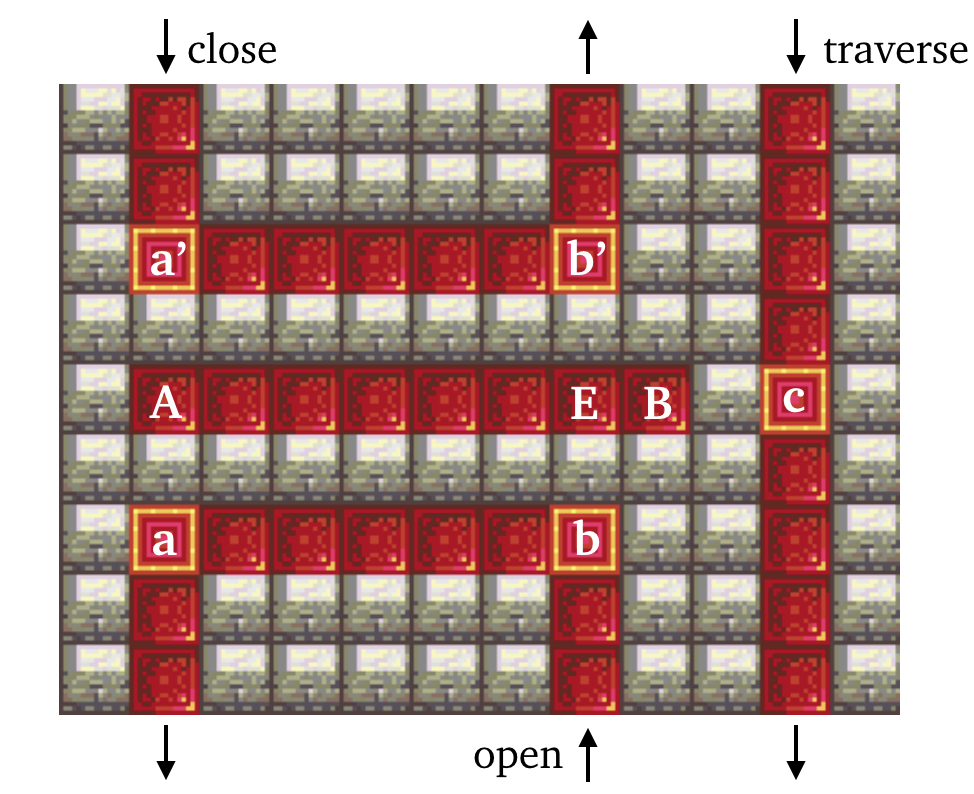}
		\caption{Door gadget. The red carpet tiles are floor tiles, and the stone tiles are castle walls. The tiles with yellow rim are safe tiles.}\label{fig_pspace_door}
	\end{figure}
	
	As shown in Figure~\ref{fig_pspace_door}, in the middle of the door gadget there is a horizontal room of length $d+2$.
	In the room at the tile labeled by $E$ we place an Enemy Door Gadget Unit, who is an Archer. He has attack range $2$, and can deal $1$ damage to player Lord. On the other hand, the player Lord cannot damage him at all.
	If the Lord goes along the traverse path, her safe tile will be position $c$. Then if the enemy Archer is at position $E$, he will move to position $B$, so that the player gets one damage from the Archer.
	
	To open the door, the Lord goes through the path below the room, from right to left. Positions $b$ and $a$ are safe tiles. When she stays on position $b$, no matter where the enemy unit is currently located in the room, the enemy unit can always shoot her from position $E$. When she stays on position $a$, the enemy unit moves to position $A$ and shoots her again. Then the Lord moves away, and the enemy unit stays at position $A$. The Lord has lost $2$ HP. After opening the door, we will place a healing unit on the Lord's path to heal her back to $3$ HP.
	
	Before going through the traverse path, the Lord should have been already damaged by $2$ HP, and has only $1$ HP left when she goes along the path to the right of the room. So if the enemy unit is not on position $A$, then he can shoot the Lord by moving to position $B$; otherwise he will not hurt the Lord because the Lord is out of range. If the Lord is shot by the enemy unit, she dies and the game is over. After traversing pass the enemy unit room, we place a healing unit on the Lord's path to heal her back to $3$ HP.
	
	To close the door, the Lord goes along the close path above the room from left to right, which is symmetric to the open path. When she stands on position $b'$, the enemy unit inside the room is lured back to position $E$, and will stay on position $E$ afterwards. After closing the door, we place a healing unit on the Lord's path to heal her back to $3$ HP.
	
	\black{Crossover gadget}
	
	When two paths cross each other, we always let one path be horizontal and the other be vertical.
	Relative to the whole map, each tile has a horizontal coordinate and a vertical coordinate.
	We arrange the crossings carefully to make sure that each tile at the crossing of two paths has both vertical and horizontal coordinates modulo $d$ equal $0$.
	Around the crossing tile, on the horizontal path, the safe tiles' distance from the crossing modulo $d$ equals $s_1$, while  on the vertical path, the safe tiles' distance from the crossing modulo $d$ equals $s_2$, satisfying:
	\begin{compactitem}
		\item $s_1 \neq s_2$, and
		\item $s_1 \neq d - s_2$, and
		\item $d - s_1 \neq s_2$, and
		\item $d - s_1 \neq d - s_2$.
	\end{compactitem}
	
	Also, before and after each crossing, there are straight paths long enough to hide powerful Snipers behind the walls to shoot the Lord if she is not in a safe tile. So the Lord can only stay in the safe tiles.
	If the Lord was moving right, but at the crossing she chose to move down (or up), then because she must stay at the tiles of distance $i \cdot d + s_1$ for integer $i$ from the crossing, but the safe tiles are of distance $i \cdot d + s_2$ (or $i \cdot d - s_2$) for integer $i$ from the crossing, so the Lord will be shot by Snipers behind the walls. Similarly, if the Lord was moving left, down or up, she cannot turn into any other direction at the crossing.
	
\begin{figure}
		\centering
		\includegraphics[width = .7\textwidth]{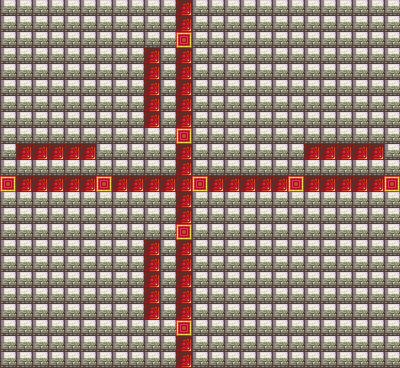}
		\caption{Crossover gadget. The tiles with yellow rim are safe tiles.}\label{fig_pspace_crossing}
	\end{figure}

	Figure~\ref{fig_pspace_crossing} shows an crossover gadget. Here $d = 6$, $s_1 = 1$ and $s_2 = 3$. When the Lord goes from left to right, the first tile after the crossing is the safe tile. When she goes top-down, the third tile after the crossing is the safe tile. The rooms in the walls are sniper rooms, where the Enemy Snipers Units would shoot the Lord if she does not stay on the safe tiles.
	
	\begin{figure}
		\centering
		\begin{subfigure}[b]{0.4\linewidth}
			\centering
			\includegraphics[scale = 2]{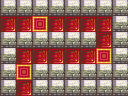}
			\caption{Turning from right to down/from up to left.}
		\end{subfigure}
		\begin{subfigure}[b]{0.4\linewidth}
			\centering
			\includegraphics[scale = 2]{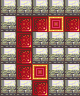}
			\caption{Turning from left to down/from up to right.}
		\end{subfigure}
		\begin{subfigure}[b]{0.4\linewidth}
			\centering
			\includegraphics[scale = 2]{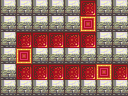}
			\caption{Turning from down to right/from left to up.}
		\end{subfigure}
		\begin{subfigure}[b]{0.4\linewidth}
			\centering
			\includegraphics[scale = 2]{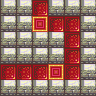}
			\caption{Turning from down to left/from right to up.}
		\end{subfigure}
		\caption{Turning gadgets. The tiles with yellow rim are safe tiles.}\label{fig_pspace_turn}
	\end{figure}
	
	\black{Turning gadget}
	
	The turning gadgets are question mark shaped curves that shifts the positions of safe tiles between horizontal paths and vertical paths, so that the coordinates of the safe tiles satisfy the condition described in the crossover gadget. Figure~\ref{fig_pspace_turn} shows the turning gadgets of all directions.
	
	\black{Healing and damaging units used in door gadgets}
	
	A Healing Unit is a player Cleric unit in a single-tile room, using the Psysic Staff whose healing range is from $2$ to $10$. The distance from the room to the nearest tile, which is a safe tile, is $2$. So when the Lord moves to the safe tile, the Cleric can heal the Lord to $3$ HP.
	
	A Damaging Unit is a enemy unit of attack range $2$ in a similar single-tile room. He can deal $2$ damage to the Lord. We also give him \emph{Def} no less than the Lord's \emph{Atk} so that he won't take any damage from the Lord.
	
\begin{figure}
		\centering
		\includegraphics[scale=2.5]{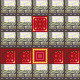}
		\caption{Healing and damaging unit. When the Lord stays on the yellow rim tile, she gets healed or damaged by the unit in the room.}\label{fig_pspace_healing}
	\end{figure}

	Figure~\ref{fig_pspace_healing} shows the single-tile room and the main path.
	
	In this way, we have created open-close door gadgets and crossover gadgets. This means Simplified FE can be reduced from TQBF. Therefore Simplified FE is PSPACE-complete.
	
	This construction can be applied to other TRPGs, as long as they support distance $2$ attacks and distance $2$ healing, and have impatient enemy units that takes the initiative to chase the player unit down.
	
	\subsection{Constructing one-way paths using two-way doors} \label{sec_oneway}
	
	This section explains the first two points in Claim~\ref{claim_oneway}, whose conditions are satisfied by our construction of the open-close door gadgets.
	
	Viglietta's reduction from TQBF \cite{viglietta2014gaming} uses open-close door gadgets to construct clause gadgets and quantifier gadgets for TQBF. In those gadgets, there are one-way paths containing sequences of open paths, close paths and traverse paths of different door gadgets. The open paths and close paths are supposed to be one-way. In our construction, we do not have one-way paths. In the door gadgets we have created, the open paths and close paths can be traversed in both directions. By going through an open path reversely, the door will be closed. By going through a close path reversely, the door will be open. It is like a ``switch'' that can be turned on and off by pushing the bar to the two sides.
	
\begin{figure}
		\centering
		\includegraphics[scale=0.9]{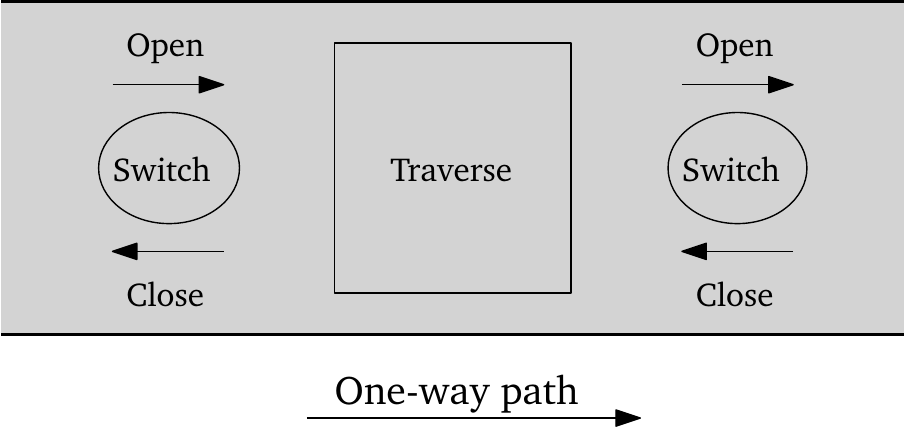}
		\caption{Using two-direction doors to create single-direction paths.}\label{fig_pspace_oneway}
	\end{figure}

	To ensure a path in a quantifier or clause gadgets to be one-directional, we add the following structure in the path, as shown in Figure~\ref{fig_pspace_oneway}. The traverse path is guarded by two switches on both sides. When going from left to right, the door is open and can be traversed. When going from right to left, the door is closed and cannot be traversed.
	
	\section{Open Problems}\label{sec_open}
	
	\black{Monotone FE}
	
	Consider Simplified FE without any healing units. The HP of each unit is always non-increasing. Is this game still PSPACE-complete, or is it easier in computational complexity?
	
	\black{Range-$1$ FE}
	
	In the proofs, we constructed a lot of units with attack range or healing range $2$. This is convenient because we can separate the map into disjoint parts, so that units cannot enter the paths of other units. Thus, a interesting question would be: can we prove the game is PSPACE-complete even when each unit has either attack range $1$ or healing range $1$?
	
	\section*{Acknowledgments}
	Last year my advisor Russell Impagliazzo gave an Algorithms course, and I was one of the TAs.
	The final project of the course was to choose a game and study its algorithms/complexity.
	I have graded a lot of students' submissions, which are very interesting and insightful.
	That gives the motivation to study the complexity of Fire Emblem, my favorite video game series.
	
	I would like to thank my friend Shiwei Weng, who provided some helpful comments on a previous draft.
	I am also thankful to the awesome map drawing tool from Serenes Forest, uploaded by user BwdYeti \cite{Serenesforest}.
	Finally I would like to thank the game developers of INTELLIGENT SYSTEMS, for making this wonderful series.

\bibliographystyle{plain}
\bibliography{refs}
\end{document}